\documentclass[10pt,a4paper,twocolumn]{article}
\usepackage[utf8]{inputenc}
\usepackage[T1]{fontenc}
\usepackage[a4paper,margin=2.5cm]{geometry}
\usepackage{amsmath,amssymb,amsfonts,bm}
\usepackage{graphicx}
\usepackage{textcomp}
\usepackage{xcolor}
\usepackage{float}
\usepackage{caption}
\usepackage[colorlinks=true,linkcolor=blue,citecolor=blue,urlcolor=blue]{hyperref}
\usepackage{xurl}
\usepackage{csquotes}
\usepackage{authblk}

\usepackage{tikz}
\usetikzlibrary{shapes, arrows.meta, positioning}

\usepackage[backend=biber, style=ieee, isbn=false, sortcites, maxbibnames=6, minbibnames=1]{biblatex}
\title{\textbf{On the Practical Feasibility of Harvest-Now, Decrypt-Later Attacks}}

\author[1]{Javier Blanco-Romero}
\author[1]{Florina Almenares Mendoza}
\author[1]{Carlos Garc\'ia Rubio}
\author[1]{Celeste Campo}
\author[1]{Daniel D\'iaz S\'anchez}

\affil[1]{Departamento de Ingenier\'ia Telem\'atica, Universidad Carlos III de Madrid, Madrid, Spain}

\begin{document}

\maketitle
		
\begin{abstract}
Harvest-now, decrypt-later (HN-DL) attacks threaten today's encrypted communications by archiving ciphertext until a quantum computer can break the underlying key exchange. This paper reframes HN-DL as an economic problem, quantifying adversary costs across Transport Layer Security (TLS) 1.2, TLS 1.3, QUIC, and Secure Shell (SSH) with an open-source testbed that reproduces the full attack sequence. Our model shows that retaining intercepted traffic is economically trivial, shifting the defensive question from whether an adversary can archive to how much decryption will cost. We evaluate protocol configuration strategies that act along two independent cost axes: storage overhead and quantum workload. Beyond the ongoing migration to post-quantum cryptography, these strategies provide defense in depth with current infrastructure. Encrypted Client Hello forces indiscriminate bulk collection, inflating the archive the adversary must retain, while aggressive rekeying and larger key exchange parameters multiply the quantum computations required to recover plaintext. Because storage inflation penalizes both sides while quantum cost inflation targets the adversary alone, rekeying and key size selection offer the strongest defensive levers.
\end{abstract}


%% ====================================================================
%%  Section 1 — Introduction
%% ====================================================================
\section{Introduction}
\label{sec:introduction}

In a \textit{harvest-now, decrypt-later} (HN-DL) attack, an adversary records encrypted communications today and decrypts them once a quantum computer can break the underlying key exchange. The threat is most acute for long-lived sensitive data and is shaped by the presence or absence of \textit{forward secrecy}: the property that compromise of a server's long-term key does not expose past session keys. Modern protocols achieve forward secrecy through \textit{ephemeral key exchange}, but older modes and certain resumption paths lack this protection. Large-scale interception of encrypted traffic by state actors is well documented~\cite{pclob2023section702,prg2013liberty,landau2013snowden}, and the accumulating archive grows more valuable as quantum hardware matures~\cite{NIST-hndl,kagai2025harvest}.

The quantum threat to public-key cryptography makes this urgent. Shor's algorithm renders RSA and ECC key exchanges polynomial-time solvable, while symmetric ciphers remain secure at doubled key sizes; the adversary therefore targets the key exchange, not the bulk cipher.

The threat has moved beyond academic interest. The National Institute of Standards and Technology (NIST) finalized its first three post-quantum cryptography (PQC) standards in August~2024~\cite{nist2024fips203}, and major infrastructure providers are already deploying hybrid key exchange in production. As of late~2025, over half of human-initiated traffic on the Cloudflare network uses post-quantum key agreement~\cite{cloudflare2025pq}, and Google has integrated PQC across Chrome and its internal services~\cite{google2026quantum}. Amazon Web Services (AWS) offers hybrid post-quantum Transport Layer Security (TLS) on its Application and Network Load Balancers~\cite{aws2025pqtls}, while Microsoft's Quantum Safe Program targets full ecosystem transition by~2033~\cite{microsoft2025qsp}. Apple's PQ3 protocol introduced post-quantum rekeying for iMessage, achieving what Apple terms ``Level~3'' messaging security~\cite{apple2024pq3}. This industry mobilization reflects a consensus that the harvest window is open and every day of delay enlarges the corpus of quantum-vulnerable ciphertext.

The practical feasibility of HN-DL depends not only on cryptographic weakness but also on storage economics. Petabyte-scale archival costs have collapsed by roughly 95\% since 2010, though AI-driven demand has introduced renewed price pressure on semiconductor storage media (Section~\ref{sec:cost-analysis}). State-level storage capacity continues to grow. Combined with a single interconnection provider reporting 68~exabytes of data traffic globally in 2024~\cite{decix2025traffic}, the volume of available ciphertext is vast. Gaining sustained access to that traffic, whether through backbone fiber taps, Internet Service Provider (ISP) cooperation, or data-center presence, remains a substantial operational prerequisite that our model does not quantify. What we do quantify is the downstream retention problem: once interception capability exists, storing the resulting archive is well within economic reach. We therefore reframe HN-DL as an economic problem, isolating the adversary's data retention costs and evaluating how existing protocol mechanisms can inflate them.

We offer three principal advances. First, we provide an open-source simulation pipeline~\cite{perlabuc3m_hndl} reproducing the HN-DL sequence against TLS~1.2, TLS~1.3, QUIC, and Secure Shell (SSH) (Section~\ref{sec:protocol-analysis}). Second, we quantify mass surveillance storage economics by deriving a per-protocol overhead ratio that maps retention requirements to state intelligence budgets (Section~\ref{sec:cost-analysis}). Third, we assess configurable defense in depth strategies that inflate the adversary's quantum workload. SSH rekeying and larger key exchange parameters multiply the required quantum computations with negligible bandwidth overhead, while Encrypted Client Hello (ECH) degrades metadata triage. We bound the rekeying multiplier analytically and identify the absence of in-band ephemeral rekeying in TLS~1.3 and QUIC as a critical protocol gap (Section~\ref{sec:mitigation}).

%% ====================================================================
%%  Section 2 — Related Work
%% ====================================================================
\section{Related Work}
\label{sec:related-work}

The HN-DL threat sits at the intersection of quantum cryptanalysis, protocol security, traffic analysis, and surveillance economics. This section positions the present work against these bodies of literature.

\subsection{Quantum Threat and HN-DL Risk Frameworks}

The quantum threat to public-key cryptography is well established. Shor's algorithm solves integer factorization and discrete logarithms in polynomial time on a cryptographically relevant quantum computer (CRQC), breaking RSA, DH, and ECC~\cite{chen2016report}. Because symmetric ciphers survive at doubled key sizes, HN-DL economics reduce to the key exchange problem.

Mosca's inequality~\cite{mosca2018cybersecurity} (Section~\ref{sec:quantum-threat}) frames the migration urgency as a race between data shelf-life, migration time, and quantum arrival. The framework is influential but deliberately abstract: it provides no mechanism for quantifying the adversary's operational cost during the harvest phase, nor does it distinguish between protocols that offer per-session isolation and those that collapse entire corpora under a single key. Kagai~\cite{kagai2025harvest} extended this model by formalizing HN-DL as a temporal cyberweapon, but both works treat HN-DL as a binary event (vulnerable or not) rather than a graduated economic decision.

NIST IR~8105~\cite{chen2016report} catalogues the quantum threat to standard cryptographic primitives and recommends migration timelines, but does not model the adversary's storage or computational cost. The European Telecommunications Standards Institute (ETSI) explicitly identified the threat model of capturing internet traffic in an encrypted state for later quantum decryption as a near-term concern in its quantum-safe cryptography white paper~\cite{etsi2015quantumsafe}, elevating the risk from an academic concern to a recognized standards-body threat category. Expert surveys by Mosca and Piani~\cite{mosca2024gri} provide the timeline input that the Mosca inequality requires but do not themselves supply. The present work fills a complementary gap by introducing a per-session storage accounting ($\alpha$) and a quantum computational cost metric ($E$) to evaluate the economic burden, validating both experimentally and scaling them to global traffic volumes.

\subsection{Protocol Defense and Forward Secrecy}

Formal analysis establishes the security properties achieved by modern protocols under classical assumptions. TLS~1.3 now dominates deployment, with Cloudflare reporting over 93\% of connections~\cite{cloudflare2024pq}, while mobile traffic splits between TLS~1.3 (52\%) and QUIC (45\%)~\cite{jimenez2025parrot}. Cremers et~al.~\cite{cremers2017comprehensive} provided the first comprehensive symbolic analysis of TLS~1.3, while Dowling et~al.~\cite{dowling2021cryptographic} contributed a computational proof in the multi-stage key exchange model. For SSH, analogous computational and post-quantum security proofs confirm the robustness of its transport layer~\cite{bergsma2014multi,benvcina2025post}. However, these analyses assume the underlying key exchange mechanism remains hard, which is precisely the assumption HN-DL exploits. 

Prior work on defending these protocols against cryptanalytic progress and traffic analysis divides into volumetric and cryptographic approaches. Record padding, studied extensively for website fingerprinting~\cite{dyer2012peek,cai2014systematic,cherubin2017website}, provides a measurable increase in the adversary's storage burden. Yet it remains inherently inefficient because it degrades defender bandwidth concurrently and leaves the entire session vulnerable to a single quantum key compromise. Periodic cryptographic rekeying operates on a different axis entirely. SSH's rekeying mechanism institutes a strict forward-secrecy boundary that prevents compromise cascades~\cite{bergsma2014multi,di1999forget}, and the IETF is now standardizing analogous in-band rekeying for TLS~1.3~\cite{ietf-tls-extended-key-update-09} and QUIC~\cite{ietf-quic-extended-key-update-02}. Our work extends this line to the quantum setting: each independent key exchange forces a separate run of Shor's algorithm, transforming the adversary's scarcest resource into a per-session cost bound (Section~\ref{sec:mit-quantum-cost}).

Post-quantum handshake performance has been extensively benchmarked~\cite{paquin2020benchmarking,schwabe2020post}, and formal verification of hybrid protocol transitions confirms the maturation of the destination state~\cite{bhargavan2024formal,linker2025formal}. While these migration studies focus on the post-quantum destination, structural defenses such as rekeying and key size selection retain value as defense in depth: should a standardized PQC algorithm prove vulnerable, they bound the resulting exposure window. Our work quantifies what it costs an adversary to violate current protocols retroactively (Section~\ref{sec:cost-analysis}) and establishes what defenders can do today to raise that barrier (Section~\ref{sec:mitigation}).

\subsection{Surveillance Economics and Storage Cost Modeling}

The economics of mass surveillance have received attention from social-science, policy, and intelligence-oversight perspectives. The PCLOB oversight report~\cite{pclob2023section702} documented the operational scale of U.S.\ collection programmes, and Danezis and Wittneben~\cite{danezis2006economics} modelled surveillance and target selection using social network theory. On the metadata axis, the ENISA report on encrypted traffic analysis~\cite{enisa2019eta} catalogued techniques (application fingerprinting, website fingerprinting, and entropy-based content typing) that allow an adversary to triage and prioritize archived sessions without decryption. None of these works develops a per-protocol, per-session cost metric suitable for quantitative comparison across protocol variants.

In the storage domain, per-byte storage costs have declined steeply over recent decades~\cite{nsa_utah_capacity}, and cloud-archive pricing provides a contemporary baseline for large-scale storage~\cite{azure2026blobpricing}. These sources provide the cost inputs but no mapping to protocol-level overhead. The protocol overhead ratio $\alpha$ used in this work (Section~\ref{sec:cost-analysis}) fills this gap: it maps protocol design choices (handshake size, framing overhead, padding) to a single dimensionless quantity that multiplies the adversary's raw storage cost.

%% ====================================================================
%%  Section 3 — Background and Threat Model
%% ====================================================================
\section{Threat Model}
\label{sec:background}

The archetypal HN-DL adversary is a state-level actor whose operational profile has three elements. First, large-scale interception: historical programs like Section~215~\cite{pclob2014section215} established the precedent for indiscriminate bulk data collection. Modern programs record encrypted internet traffic at volume through passive backbone fiber tapping and server-side acquisition, generating no alerts and leaving no forensic trace~\cite{pclob2023section702,prg2013liberty,landau2013snowden}. Second, persistent storage: tiered architectures sustain multi-decade retention at marginal per-gigabyte cost (Section~\ref{sec:cost-analysis} quantifies this in detail). Third, deferred quantum cryptanalysis: the adversary waits for a CRQC, then processes the archive. Both U.S.\ and European national security guidelines acknowledge this threat model: the ANSSI/BSI/NLNCSA position paper explicitly names the ``store-now-decrypt-later'' scenario~\cite{anssi2024qkd}, and the NSA's CNSA~2.0 advisory mandates urgent post-quantum migration on timelines consistent with this threat~\cite{nsa2022cnsa2} (Section~\ref{sec:discussion} discusses the implications).

\label{sec:quantum-threat}
The feasibility of this deferred attack depends on when quantum hardware arrives. Mosca~\cite{mosca2018cybersecurity} formalized the migration urgency as:
\begin{equation}
\label{eq:mosca}
x + y > z,
\end{equation}
where $x$ is the data shelf-life, $y$ is the migration time to quantum-resistant algorithms, and $z$ is the time until Q-Day. Because $y$ can exceed a decade for complex infrastructures~\cite{chen2016report}, the urgency for migration is immediate even if $z$ lies a decade away. Expert surveys place Q-Day in the 2030--2040 window, with roughly 50\% probability of a CRQC breaking RSA-2048 within 15~years~\cite{mosca2024gri,kagai2025harvest,chen2016report}.

%% ====================================================================
%%  Section 4 — Protocol Analysis
%% ====================================================================
\section{Protocol Analysis}
\label{sec:protocol-analysis}

Forward secrecy (Section~\ref{sec:introduction}) partitions protocols into two HN-DL categories. Without it (e.g., TLS~1.2 RSA key transport), compromising the server's long-term key unlocks every past session. Forward-secret protocols (TLS~1.3, SSH, TLS~1.2 with DHE/ECDHE) resist this classical collapse but remain quantum-vulnerable: the adversary must solve the ephemeral discrete logarithm for each targeted session. Forward secrecy therefore raises the attack cost from a single key compromise to per-session cryptanalysis, but does not eliminate the quantum threat.

This section analyzes TLS~1.2, TLS~1.3, QUIC, and SSH based on these cryptographic properties, then validates each attack path with an end-to-end simulation.
\subsection{Testbed and Methodology}
\label{sec:exp-testbed}

A Python orchestrator chains protocol-specific capture and derivation modules. The testbed uses patched OpenSSL~3.6.0 and OpenSSH~9.9p2 for traffic generation, tshark ($\geq$4.0) for packet capture, and Python ($\geq$3.10) for orchestration and key derivation. All traffic flows over the loopback interface, eliminating packet loss, TCP retransmissions, and MTU fragmentation overhead. Consequently, every measured byte count represents a strict lower bound on the real-world adversary's storage burden.

To simulate quantum key recovery, source-level patches log the ephemeral private keys a CRQC would obtain by solving the ECDLP. Each derivation module must reconstruct all session secrets exclusively from the PCAP and this simulated quantum output. We verify the reconstruction in two layers: a byte-for-byte comparison against the ground-truth keylog (captured via the standard \texttt{SSLKEYLOGFILE} environment variable for TLS/QUIC and internal state extraction for SSH), followed by tshark decryption of the application data to prove full compromise.

\subsection{TLS~1.2}
\label{sec:tls12-hndl}

\subsubsection*{Key Exchange}
TLS~1.2~\cite{rfc5246} derives session keys from a pre-master secret via the TLS Pseudorandom Function (PRF). Vulnerability to HN-DL depends entirely on the key-exchange family. In \emph{RSA key transport}, the client encrypts a random pre-master secret under the server's RSA public key; because no ephemeral material is involved, obtaining the private key decrypts every session that ever used it. In \emph{ephemeral DH} (DHE/ECDHE), fresh one-time key pairs provide forward secrecy against classical adversaries, but a CRQC can solve the discrete logarithm for each recorded exchange. TLS~1.3 removed all non-forward-secret key exchanges by design~\cite{rfc8446}.

\subsubsection*{Testbed Attack}
We target the static RSA key transport mode, where no ephemeral exchange protects the session. The Python derivation module extracts the encrypted pre-master secret from the PCAP, decrypts it via PKCS\#1~v1.5, and derives all session keys through the TLS~1.2~PRF. The derived master secret matches the \texttt{SSLKEYLOGFILE} ground truth in every run, and tshark decrypts the payload, confirming full retrospective decryption from a single compromised key.

\subsection{TLS~1.3}
\label{sec:tls13-hndl}

\subsubsection*{Key Schedule}
TLS~1.3~\cite{rfc8446} eliminates static RSA key transport; the server's certificate key is used solely for authentication, never for key encapsulation. The entire key schedule relies on chained HMAC-based Extract-and-Expand Key Derivation Function (HKDF) operations~\cite{rfc8446,rfc5869}: EarlySecret incorporates a PSK (or zero), HandshakeSecret mixes in the ephemeral shared secret~$Z$, and MasterSecret yields the application traffic keys (Appendix~\ref{appendix:tls13-singlehndl}). Because every step is deterministic once $Z$ and the plaintext transcript are known, the primary HN-DL vector is the recovery of~$Z$ via a CRQC.

This deterministic nature extends through the protocol's lifecycle features. \texttt{KeyUpdate} derives each new traffic secret from its predecessor via a single HKDF expansion~\cite{rfc8446}; because no new randomness enters the derivation, recovering the initial~$Z$ from the handshake exposes every subsequent symmetric epoch. Similarly, after a full handshake, peers may derive per-ticket PSKs from the Resumption Master Secret (Appendix~\ref{appendix:tls13-resumption}). Resumed connections proceed either in PSK-DHE mode (which introduces a fresh ECDHE exchange, preserving forward secrecy) or pure PSK mode (which enables 0-RTT data with no new exchange). Consequently, breaking one session's ECDHE not only compromises its immediate traffic but transitively exposes all subsequent 0-RTT and pure-PSK sessions relying on its cascade of resumption tickets.

\subsubsection*{Testbed Attack}
For 1-RTT connections, the derivation module computes $Z$ from a simulated-quantum ephemeral key and executes the full RFC~8446 key schedule. All four traffic secrets correctly match the ground truth. For 0-RTT resumption, the pipeline successfully recovers the PSK from a prior session's derived Resumption Master Secret and generates the early traffic secret for the resumed session. The result exactly matches the ground truth, empirically confirming that a single ECDHE compromise unravels the entire pure-PSK resumption chain.

\subsection{QUIC}
\label{sec:quic-hndl}

\subsubsection*{Key Exchange}
QUIC~\cite{rfc9000} carries the TLS~1.3 handshake inside its own transport layer~\cite{rfc9001}. The \texttt{ClientHello} and \texttt{ServerHello} travel in QUIC Initial packets (encrypted with keys derived from the observable Destination Connection~ID), while subsequent handshake messages travel in QUIC Handshake packets (encrypted with the handshake traffic secrets). Because the underlying key schedule is identical to TLS~1.3, the HN-DL vulnerability profile is unchanged: recovery of a single ephemeral private key suffices to reconstruct all session secrets. At the transport layer, UDP framing and cleartext connection~IDs produce small constant-factor adjustments in storage overhead (Section~\ref{sec:cost-analysis}) but do not alter the core cryptographic vulnerability.

\subsubsection*{Testbed Attack}
A patched OpenSSL demo ($\approx$80~lines) links against our instrumented library and provides a QUIC server with keylog support. The Python module decrypts the QUIC Initial packets natively by deriving Initial keys from the Destination Connection~ID per RFC~9001. Once the handshake payloads are recovered, the module computes $Z = \text{X25519}(a, g^b)$ and executes the standard TLS~1.3 key schedule (Appendix~\ref{appendix:tls13-singlehndl}). The pipeline successfully decrypts the remaining handshake and application packets, and all four derived traffic secrets match the \texttt{SSLKEYLOGFILE} ground truth. This confirms that QUIC's transport-layer encryption of the handshake provides no supplementary protection against HN-DL; once the ephemeral DH problem is solved, the QUIC packet protection keys fall identically to TLS~1.3.

\subsection{SSH}
\label{sec:ssh-hndl}

\subsubsection*{Key Exchange}
SSH~\cite{rfc4251,rfc4253} derives forward secrecy by design: each session performs an ephemeral ECDH exchange for the shared secret~$K$, while the host key serves strictly for authentication. The IETF has further deprecated the weakest RSA-based SSH key exchange methods~\cite{rfc9142}. The derivation is a single stage: the exchange hash~$H$ binds the handshake transcript, and all six session keys follow directly~\cite{rfc4253}. Consequently, once $K$ is known, all keys are immediately derivable. A CRQC solves the ECDLP from the captured ephemeral public keys, recovers~$K$, and computes the full symmetric hierarchy.

Unlike TLS~1.3's deterministic \texttt{KeyUpdate} chain, SSH supports true cryptographic rekeying. A rekey event initiates a full new Diffie-Hellman exchange, producing a fresh shared secret $K'$ and exchange hash $H'$ while retaining the original session identifier~\cite{rfc4253}. Because fresh ephemeral randomness is injected, each rekey epoch must be broken independently by the adversary. This isolated epoch property is exploited directly as a volumetric mitigation in Section~\ref{sec:mit-rekey}.

\subsubsection*{Testbed Attack}
We target the \texttt{curve25519-sha256} exchange method (the OpenSSH default). Unlike TLS, SSH lacks a standardized `keylogfile' export. Therefore, our ground truth relies on source-level instrumentation of the OpenSSH service daemon to explicitly export the active ephemeral secrets to a secured file during the handshake. Given the simulated-quantum derivation of the client's ephemeral key, the Python derivation module computes $K = \text{X25519}(a, g^b)$ and applies the RFC~4253 key derivation. All six derived symmetric keys match the instrumented ground truth, validating the full capture-to-decrypt pipeline.

\subsection{Comparative Taxonomy}
\label{sec:comparative-taxonomy}

Table~\ref{tab:protocol-taxonomy} synthesizes the vulnerability profile of each protocol and mode. We successfully validated every attack path end-to-end; recovering the underlying secret (RSA private key for TLS~1.2 RSA, ephemeral $Z$ for TLS~1.3 and QUIC, $Z$ plus the PSK chain for 0-RTT, ephemeral $K$ for SSH) deterministically yields the plaintext.

\begin{table}[htbp]
\centering
\small
\caption{HN-DL vulnerability taxonomy across protocols and modes.}
\label{tab:protocol-taxonomy}
\resizebox{\columnwidth}{!}{%
\begin{tabular}{lcccc}
\hline
\textbf{Protocol / Mode} & \textbf{Forward} & \textbf{Classical} & \textbf{Quantum} & \textbf{Recovery} \\
 & \textbf{secrecy} & \textbf{HN-DL} & \textbf{HN-DL} & \textbf{scope} \\
\hline
TLS 1.2 (RSA)       & No  & Vulnerable  & Vulnerable & All sessions/key \\
TLS 1.2 (DHE/ECDHE) & Yes & Resistant   & Vulnerable & Per session \\
TLS 1.3 (1-RTT)     & Yes & Resistant   & Vulnerable & Per session \\
TLS 1.3 (KeyUpdate) & Yes & Resistant   & Vulnerable & Per session$^\dagger$ \\
TLS 1.3 (PSK-DHE)   & Yes & Resistant   & Vulnerable & Per session \\
TLS 1.3 (0-RTT)     & No  & Vulnerable  & Vulnerable & Per PSK chain (0-RTT only) \\
QUIC (ECDHE)         & Yes & Resistant   & Vulnerable & Per session \\
SSH (ECDH)           & Yes & Resistant   & Vulnerable & Per exchange \\
\hline
\multicolumn{5}{l}{\footnotesize $^\dagger$ Deterministic chain: breaking initial ECDHE exposes all KeyUpdate epochs.}
\end{tabular}}
\end{table}

The critical variable is recovery scope. TLS~1.3's KeyUpdate chain and PSK resumption cascade widen the blast radius of a single ECDHE compromise beyond its originating session, whereas SSH's independent rekey exchanges confine exposure to a single epoch. Sections~\ref{sec:cost-analysis} and~\ref{sec:mitigation} exploit this distinction to quantify storage costs and show how configuration changes inflate the adversary's budget.

%% ====================================================================
%%  Section 5 — Cost Analysis
%% ====================================================================
\section{Cost Analysis}
\label{sec:cost-analysis}

The previous sections establish that retrospective decryption is deterministic once the ephemeral key is recovered.
The remaining question is economic: can an adversary afford to maintain the resulting archives until quantum computers arrive?
We assess only the at-rest storage burden. Gaining sustained access to encrypted traffic requires positioning at the right network point, whether through optical fiber taps, ISP cooperation, or state-mandated data-localization infrastructure. These upstream interception costs are operationally significant and highly variable; our model deliberately excludes them to isolate the storage baseline. We construct a byte-level storage model, validate it against testbed captures, and propagate input uncertainty through Monte Carlo simulation.

\subsection{Per-Session Storage Model}
\label{sec:per-session-storage}

Because secure ciphertexts are computationally indistinguishable from uniform randomness, the Shannon entropy bound dictates they can neither be meaningfully compressed nor deduplicated. The captured byte count represents an absolute, incompressible floor on the adversary's storage budget.

To formulate the storage footprint $S(P)$ of a session carrying $P$~bytes of application plaintext, we decompose the traffic into a static initialization phase and a data transmission phase:
\begin{equation}
\label{eq:session-storage}
S(P) = H + C + \sum_{i=1}^{n}\bigl(p_i + \omega_i\bigr) + n_{\text{data}}(P)\,\ell
\end{equation}
Here, $H$ is the handshake transcript and $C$ encompasses additional session-layer setup (such as the $1{,}109$~B SSH channel multiplexing negotiation). The plaintext signal $P$ is split into $n = \lceil P/M \rceil$ discrete application records, where $M$ is the protocol's payload capacity boundary ($2^{14}$~B for TLS, $2^{15}$~B for SSH). Each record payload $p_i$ is encapsulated with a per-record framing penalty $\omega_i = r + t + q(p_i) + e$ mapping the record header~$r$, the authentication tag~$t$, any requisite alignment padding~$q(p_i)$, and structure bytes~$e$. Finally, the $n$ records are multiplexed across $n_{\text{data}}(P)$ transport-layer packets, each incurring an L2/3/4 header tax of $\ell$~bytes (54~B for TCP, 42~B for UDP).

Recognizing that $\sum_{i=1}^n p_i = P$ by construction, and abstracting the discrete record penalty to its expected mean $\omega = \langle \omega_i \rangle$, the record summation collapses into a linear upper bound:
\begin{equation}
\label{eq:session-storage-simplified}
S(P) = H + C + P + n\,\omega + n_{\text{data}}(P)\,\ell
\end{equation}
This linear model separates the core payload ($P$) from the cryptographic and transport overhead ($H, C, n\,\omega$). The expected record overhead is derived analytically from the protocol specifications~\cite{rfc8446,rfc4253}: $\omega = 22$~B exactly for TLS~1.3, and $\omega \approx 28.5$~B on average for SSH due to the uniform distribution of block-alignment padding. The transport term $n_{\text{data}}\,\ell$ fluctuates slightly with volatile network MTU limits; however, an adversary performing continuous stream reassembly prior to archiving can discard the transport headers entirely, forcing this term to vanish. Our idealised loopback captures (Section~\ref{sec:exp-testbed}) empirically isolate this deterministic cryptographic floor without random network inflation.

Finally, while initial handshake metrics dwarf the per-record framing, they are strictly fixed setup costs for the initial connection. Within our testbed, TLS~1.2 RSA requires a baseline $H \approx 1{,}620$~B, TLS~1.3 expands to $\approx 2{,}160$~B, QUIC incorporates connection~ID structures to reach $\approx 2{,}400$~B\footnote{RFC~9000 mandates that the client's first Initial datagram be padded to at least $1{,}200$\,B for path MTU discovery~\cite{rfc9000}, injecting ${\approx}700$\,B of \texttt{PADDING} frames outside the \texttt{CRYPTO} stream. Because these frames are not bound to the transcript hash, a protocol-aware adversary could strip them, reducing $H$ by ${\approx}30\%$. This saving is material only for ultra-short sessions; at scale, $\alpha_\infty$ dominates.}, and SSH demands $\geq 5{,}100$~B to satisfy its iterative negotiation sequence. This handshake footprint ($H$) cannot be pruned, compressed, or summarized by the adversary. Because the key schedules of both TLS and SSH mathematically bind the derived session keys to a cryptographic hash of the exact, byte-for-byte handshake transcript~\cite{rfc8446,rfc4253}, any attempt to strip seemingly extraneous negotiation data will corrupt the derived keys and permanently prevent retrospective decryption.

\subsection{Protocol Overhead}
\label{sec:amplification}

The \textit{protocol overhead ratio} $\alpha(P) = S(P)/P$ isolates the overhead ratio:
\begin{equation}
\label{eq:alpha-expanded}
\alpha(P) = \frac{H + C}{P} + 1 + \frac{n\,\omega}{P} + \frac{n_{\text{data}}(P)\,\ell}{P}
\end{equation}
As $P$ grows, the handshake term $(H+C)/P$ vanishes. Under stream reassembly (Section~\ref{sec:per-session-storage}), the transport term likewise vanishes, leaving the \emph{framing asymptote}
\begin{equation}
\label{eq:framing-asymptote}
\alpha_\infty = 1 + \frac{\omega}{M}
\end{equation}
Under maximum packing, $\alpha_\infty < 1.003$ for TCP-based protocols.
QUIC, constrained to ${\approx}1{,}350$~B datagrams, reaches $\alpha_\infty \approx 1.02$, reflecting the higher per-datagram framing ratio inherent to UDP transport.
An optimally efficient adversary could further tighten the archive by discarding AEAD authentication tags, stripping constant header fields, and performing stream reassembly. Appendix~\ref{appendix:minimal-archive} examines this per-record reduction for each protocol and shows that the resulting $\alpha_{\infty,\min}$ differs from $\alpha_\infty$ by less than $1.2 \times 10^{-2}$, yielding no strategically meaningful storage advantage at scale.

At the other extreme, small payloads ($P < 1$~KB) push $\alpha$ from $3.5\times$ (TLS~1.2 RSA) to $8.5\times$ (SSH), penalising the sessions most likely to carry high-value intelligence: authentication tokens, API credentials, and session keys.
Because TLS record lengths appear in the cleartext header, a rational adversary can triage by both metadata and payload size.

\begin{figure}[htbp]
\centering
\includegraphics[width=\columnwidth]{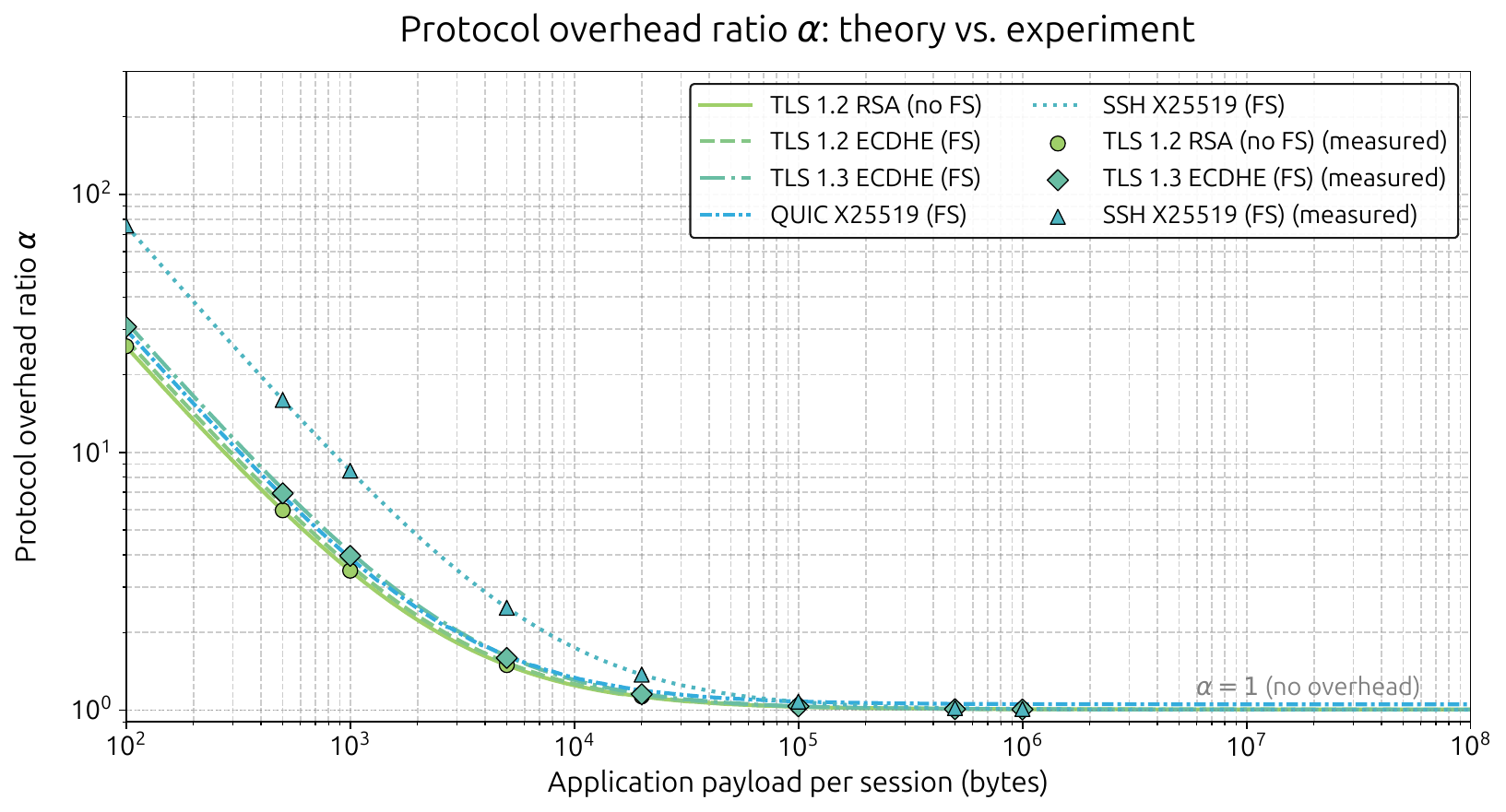}
\caption{Protocol overhead ratio~$\alpha$ vs.\ session payload.
Lines: analytical model; markers: loopback captures.
Stream reassembly lowers~$\alpha$ slightly (transport term vanishes).}
\label{fig:model-validation}
\end{figure}

Figure~\ref{fig:model-validation} validates the model against 32 loopback captures (8~payload sizes $\times$ 4~protocols, including QUIC) from the testbed of Section~\ref{sec:exp-testbed}.
Below 1\,KB, QUIC sits at $\alpha = 3.9\times$ due to UDP framing and connection-ID overhead.
Beyond 100\,KB, TCP-based protocols converge to $\alpha \approx 1.03$--$1.07$; QUIC settles near $1.05$.
Residuals satisfy $|\Delta\alpha| \leq 0.01$ at 100~B and $\leq 0.002$ beyond 1~KB.

\subsection{Storage Economics}
\label{sec:global-traffic}
\label{sec:harvesting-strategies}
\label{sec:annual-harvest-cost}

ITU estimates end user fixed+mobile broadband Internet traffic at approximately 8.8\,ZB/year in 2025~\cite{itu2025facts}. This volume corresponds to roughly $\mathcal{O}(10^{13})$ sessions per day.
At this scale, nearly all traffic falls in the regime where $\alpha \approx 1$; storage cost is strictly bounded by raw ciphertext volume.

A 1\% selective harvest requires $\mathcal{O}(10^{19})$~bytes/year.
Full capture demands $\mathcal{O}(10^{21})$~bytes/year, historically exceeding the estimated exabyte-tier capacities of facilities like the NSA Utah Data Center~\cite{nsa_utah_capacity}. However, commercial hyperscalers provide a realistic proxy for modern state capabilities: the world's largest cloud providers now operate over a thousand hyperscale data centers globally, accounting for vast infrastructure capacity~\cite{srgresearch2025hyperscale}. State actors have also mandated zettabyte-scale expansion directly, such as China's 1.8\,ZB 2025 national target~\cite{prc_miit_2023} and Russia's decentralized Yarovaya data-localization laws~\cite{yarovaya_law}.
Forward secrecy does not reduce the storage burden; it multiplies required quantum computation.
TLS~1.2 RSA requires one key recovery per server certificate, whereas TLS~1.3 and QUIC demand an independent ECDLP solution per session, and SSH requires per-session Diffie-Hellman inversion.
The adversary faces a two-dimensional cost characterized by the storage footprint $S$ and the computational effort $E \times T_q$, where $E$ is the number of ephemeral keys recovered per session and $T_q$ is the quantum computational time required per key, itself a function of the key exchange parameter size (Section~\ref{sec:mit-quantum-cost}).
Here, $E = 1$ for baseline sessions and $E = \lceil P/R \rceil$ under fresh-DH rekeying every $R$~bytes of plaintext (Section~\ref{sec:mitigation}).

Selective targeting optimises this two-dimensional cost by focusing on sectors where data retains value for decades: healthcare~\cite{bolaji2023cost}, government, and finance.
These sessions tend to be small ($P \ll 1$~MB), pushing $\alpha$ to 1.2--1.4 for TLS and up to 3.9 for interactive SSH.
A rational adversary combines bulk capture of large transfers ($\alpha \approx 1$) with selective archiving of small, high-value sessions.
Metadata surveillance is cheaper and yields significant intelligence via traffic analysis~\cite{danezis2006economics,landau2013snowden,enisa2019eta}, but it cannot replace the cryptographic plaintext recovery that HN-DL targets.

Table~\ref{tab:harvest-cost} evaluates harvest fractions against commercial cloud architectures (\$12.16--\$14.74/TB-year~\cite{aws_s3_pricing,gcp_storage_pricing}) to establish a fully-loaded, worst-case operational bound that implicitly incorporates infrastructure overheads. This models only the passive retention phase.\footnote{The capital expense of traffic interception is highly variable and tightly coupled to the adversary's operational position. Establishing targeted backbone optical taps demands significant infrastructure investment, while state-mandated data localization (e.g., Russia's Yarovaya Law~\cite{yarovaya_law}) delegates collection costs to compliant ISPs, shifting the expense rather than eliminating it. Because interception costs depend on geopolitical and infrastructural factors beyond our model's scope, we strictly isolate our quantitative bounds to the deterministic storage mechanics.}
Even under this conservative assumption, full global capture represents an $\mathcal{O}(10^{11})$~USD annual retention expense, safely within the bounds of top-tier state defense budgets. This figure accounts only for storage; the upstream interception infrastructure to actually acquire traffic at that scale constitutes a separate, substantial investment.

However, dedicated state actors operating localized tape libraries push costs significantly lower. Enterprise volume pricing for HPE LTO-9 tape establishes a pure raw media capital expenditure (CapEx) floor of just \$5.25/TB~\cite{HPE_LTO9_2024}. Modern Barium Ferrite (BaFe) magnetic tape yields a confirmed at-rest media lifespan of over 50 years with near-zero energy consumption~\cite{Fujifilm_TCO_2024}. For an adversary accumulating data over decades, this eliminates the massive recurrent capital expenditures and energy costs associated with migrating exabytes across decaying hard disk arrays. Tape provides the offline air-gapping mandated for highly classified intelligence archives. Because the media investment is a one-time capital expenditure, a state operator with deployed tape infrastructure is further insulated from the market-driven price volatility that cloud providers pass through to customers. At these rates, once an adversary has positioned itself to intercept traffic, the storage component of an HN-DL attack is economically trivial compared to the computational expense of the ``Decrypt-Later'' phase.

\begin{table}[htbp]
\centering
\caption{Annual storage cost of HN-DL archives at global scale ($\alpha \approx 1$ for bulk traffic, fully-loaded AWS operational upper bound at {\normalfont\$}12.16/TB-year). Estimates isolate at-rest retention, omitting active network interception expenses.}
\label{tab:harvest-cost}
\footnotesize
\setlength{\tabcolsep}{3.5pt}
\begin{tabular}{@{}rccc@{}}
\hline
\textbf{Harvest} & \textbf{Daily ingest} & \textbf{Annual volume} & \textbf{Annual cost} \\
\textbf{fraction} & \textbf{(PB)} & \textbf{(EB)} & \textbf{(\$/yr)} \\
\hline
1\%   &   241  &  88  & \$1.1B \\
10\%  & 2{,}411  & 880  & \$11B \\
100\% & 24{,}110 & 8{,}800 & \$107B \\
\hline
\end{tabular}
\end{table}

These point estimates rest on fixed assumptions about session size, storage pricing, and traffic volume.
To assess robustness, we propagate this uncertainty through a Monte Carlo simulation ($N = 10{,}000$ draws).
Session payload follows a log-normal distribution (median 2~MB, $\sigma_{\ln} = 1.5$), spanning from small API calls (${\sim}10$\,KB) to multi-megabyte page loads consistent with HTTP Archive observations~\cite{httparchive2024pageweight}.
Storage unit cost is drawn uniformly from $\pm 30\%$ around the fully-loaded \$12.16/TB-year cloud upper bound to account for vendor and tier variability.
For each draw, $\alpha$ is recomputed from the sampled payload via Eq.~\eqref{eq:alpha-expanded} and annual cost follows as $\mathcal{C}_1 = f \cdot V \cdot \alpha \cdot c$, where $f$ is the harvest fraction, $V = 8.8$~ZB/year, and $c$ the sampled cost.

\begin{figure*}[!t]
\centering
\includegraphics[width=\textwidth]{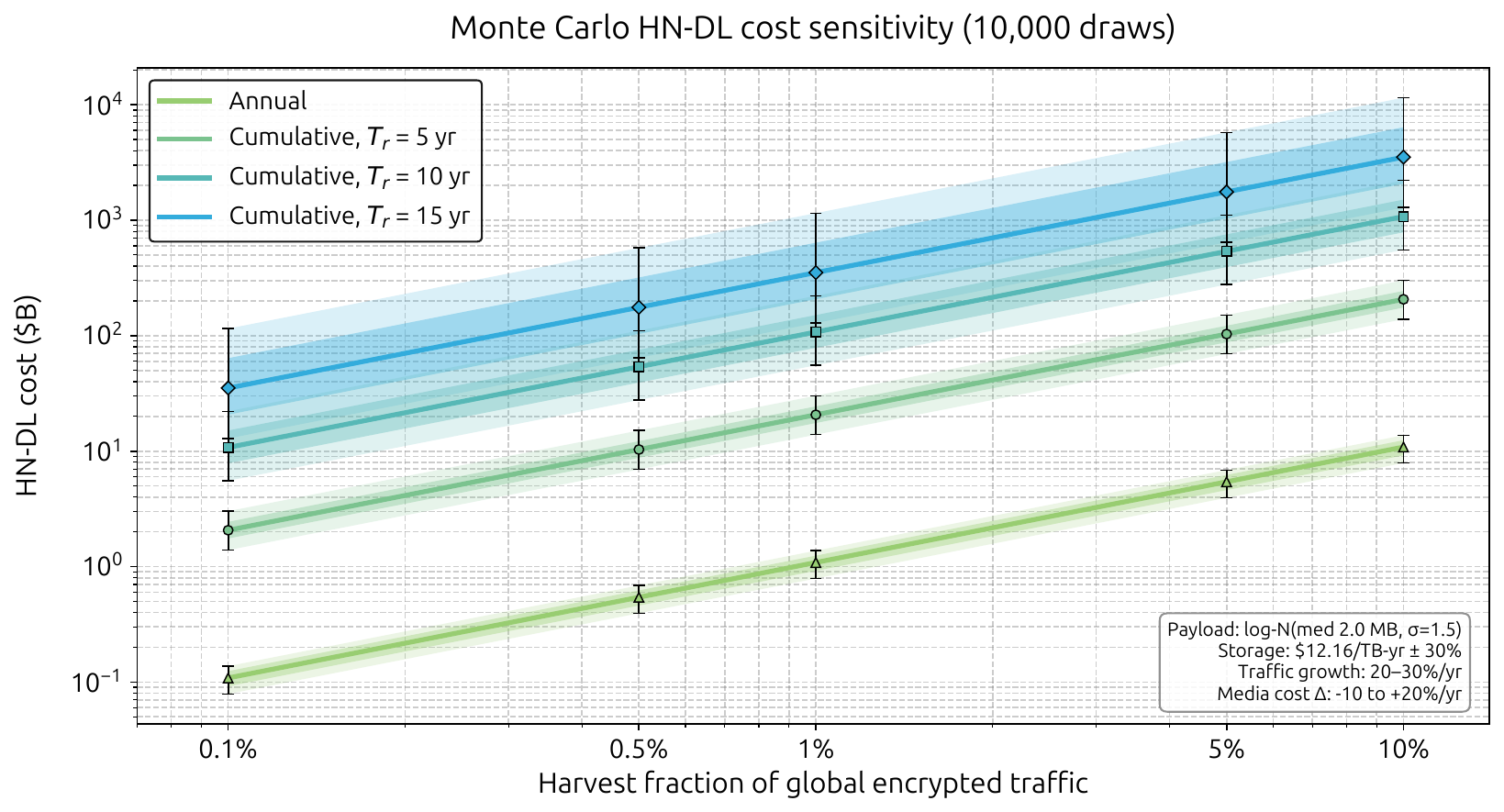}
\caption{Monte Carlo HN-DL cost as a function of harvest fraction ($10{,}000$ draws).
The plot unifies annual storage bounds (green band) with cumulative exposure over three multi-year retention horizons (green-to-blue bands, $T_r = 5, 10, 15$~years).
Solid lines mark median costs; inner bands span the 25\textsuperscript{th}--75\textsuperscript{th} percentiles and outer bands the 5\textsuperscript{th}--95\textsuperscript{th} percentiles.
The model compounds log-normal payload uncertainty (median 2~MB) and fully-loaded storage OpEx ($\pm 30\%$ around {\normalfont\$}12.16/TB-year) with compounding annual traffic growth (20--30\%/yr~\cite{cisco2020air}) and annual storage price change ($-10\%$ to $+20\%$/yr~\cite{backblaze2024costdrive,reuters2025memorychip}). Network interception costs are excluded.}
\label{fig:mc-cumulative-cost}
\end{figure*}

Figure~\ref{fig:mc-cumulative-cost} validates these scale estimates.
A 1\% selective harvest anchors confidently at $\mathcal{O}(10^{9})$~USD annually.
Storage cost uncertainty strictly drives the variance; session-size dispersion exerts only a negligible second-order effect because $\alpha \approx 1$ dominates bulk traffic physics.
Even pessimistic draws preserve the predicted scale, confirming the robustness of the $\mathcal{O}(10^9)$~USD threshold in Table~\ref{tab:harvest-cost}.

An operational program must retain captured data until decryption becomes feasible.
Expert surveys place Q-Day in the 2030--2040 window~\cite{mosca2024gri,kagai2025harvest}, mandating a total retention span $T_r \approx 5$--15~years.
The cumulative storage expenditure $\mathcal{C}(T_r)$ over this horizon compounds annual traffic growth~$\gamma$ against media-cost decline~$\delta$:
\begin{equation}
\label{eq:cumulative-cost}
\mathcal{C}(T_r) = \sum_{i=0}^{T_r-1} (T_r - i)\, V_0\, (1+\gamma)^i\, C_0\, (1-\delta)^i
\end{equation}
where $V_0$ is the initial annual volume and $C_0$ the initial unit cost. The $(T_r - i)$ factor captures the remaining retention obligation: data harvested in year~$i$ must be stored for the $T_r - i$ years until Q-Day.

Extending our Monte Carlo simulation to project long-term exposure, we draw uniform values for traffic growth ($\gamma \in [20\%, 30\%]$), consistent with the compound annual growth rates reported by Cisco~\cite{cisco2020air}. For storage cost evolution, the long-term trend favors declining media prices: HDD cost per gigabyte fell roughly 87\% between 2009 and 2022~\cite{backblaze2024costdrive}, and LTO tape capacity continues to double per generation~\cite{Fujifilm_TCO_2024}. However, AI-driven demand has sharply inflated DRAM and NAND flash prices, with secondary pressure on HDD procurement~\cite{reuters2025memorychip}; magnetic tape, produced via a distinct Barium Ferrite process, has not experienced comparable increases. We therefore model annual storage price change as $\delta \in [-10\%, +20\%]$, spanning a 10\% annual price increase through an aggressive 20\% decline. This range conservatively applies cloud pricing dynamics to all scenarios; a state operator whose hardware CapEx is already sunk faces a substantially flatter cost curve, immune to commercial price fluctuations.
Over a 10-year horizon, the aggregated cost of a 1\% selective harvest advances to $\mathcal{O}(10^{10}$--$10^{11})$~USD.
Even under this conservative framing, where cloud-grade OpEx volatility is imposed on adversaries who may never pay it, long-term accumulation presents no insurmountable economic barrier to an advanced persistent threat.

This confirms Mosca's inequality \eqref{eq:mosca}: when the required confidentiality lifetime~$x$ exceeds $z - y$ (time to quantum availability minus migration time), capture today guarantees future compromise.
For healthcare records ($x \approx 25$--50~years~\cite{bolaji2023cost}), classified communications ($x > 25$~years), and financial audits ($x > 7$~years), the archive retains strategic value well beyond~$T_r$.

%% ====================================================================
%%  Section 6 — Mitigation
%% ====================================================================
\section{Defense-in-Depth}
\label{sec:mitigation}

The adversary's cost map spans two axes: the storage burden, governed by the protocol overhead ratio~$\alpha$, and the quantum computational effort, governed by $E \times T_q$ (Section~\ref{sec:cost-analysis}). These axes are fundamentally asymmetric. Inflating~$\alpha$ requires every additional byte to traverse real network infrastructure; the defender pays active, continuous bandwidth and egress costs while the adversary absorbs the inflated volume passively at marginal cost. Inflating~$E$ or~$T_q$, by contrast, is inherently adversary-facing: each additional key exchange or larger group parameter forces fresh quantum work with negligible wire overhead. Configuration strategies that operate along the quantum axis are therefore the more cost-effective degrees of freedom.

Only post-quantum cryptography (PQC) provides fundamental security. Yet, structural friction retains value as defense in depth: should a standardized PQC algorithm fail, these layers protect the resulting HN-DL window (Section~\ref{sec:discussion}).

All measures below are server-side configurations requiring no client modification.

\subsection{Degrading Adversary Triage: Encrypted Client Hello (ECH)}
\label{sec:mit-ech}

A 1\% selective harvest of global traffic is a rounding error ($\mathcal{O}(10^9)$~USD) for a tier-one state actor, whereas absolute bulk collection imposes vast infrastructure demands (Section~\ref{sec:annual-harvest-cost}). To execute cheap selective harvesting, the adversary must identify high-value sessions (e.g., \texttt{health-records.gov}) during initial ingestion. In TLS~1.3, the primary triage selectors are the destination IP address and the cleartext Server Name Indication (SNI) extension.

Encrypted Client Hello (ECH)~\cite{ietf-tls-esni-25} mathematically conceals the true SNI by encrypting it under a provider public key distributed via DNS, exposing only a generic ``outer'' SNI on the wire. However, cryptographic cloaking is structurally insufficient if the target's IP address remains unique. ECH depends on massive coalescence: when a target hides behind a shared Content Delivery Network (CDN) or Anycast network, the adversary only sees the outer SNI and a generic load-balancer IP. The true destination uncouples from the transport metadata.

This prevents zero-cost deterministic triage. An advanced adversary must instead apply Encrypted Traffic Analysis~\cite{enisa2019eta,cai2014systematic}, using offline machine learning classifiers on packet timing and sizing metadata to probabilistically identify targets. ECH thus degrades triage into noisy, computationally expensive filtering. It forces false positives that inflate archive volumes and false negatives that miss targets. While it cannot prevent an adversary from simply archiving the entire CDN, ECH remains a defense available today to protect small-volume cohorts from algorithmic harvesting.

\subsection{Eliminating Legacy Exposure}
\label{sec:mit-legacy}

Two legacy protocol features leave sessions vulnerable to retrospective decryption and should be disabled.

\paragraph{TLS~1.2 RSA key transport.}
A single long-term RSA key compromise decrypts all sessions ever protected by that key. Disabling static RSA key exchange cipher suites removes this unbounded surface with zero performance penalty.

\paragraph{TLS~1.3 0-RTT early data.}
0-RTT data is encrypted under a PSK with no fresh ephemeral exchange. The compromise of the originating session cascades deterministically to all subsequent 0-RTT payloads. Disabling it (\texttt{ssl\_early\_data off}) eliminates the exposure at the accepted cost of one handshake round trip.

\paragraph{Bounding PSK resumption cascades.}
TLS~1.3 PSK resumption creates a transitive vulnerability: breaking one session's ECDHE exposes all PSK-derived descendants (Section~\ref{sec:tls13-hndl}). Short ticket lifetimes (e.g., 300~s instead of 86\,400~s) limit temporal exposure. Enforcing \texttt{psk\_dhe\_ke} mode ensures each resumed session introduces independent ECDHE (Section~\ref{sec:mit-quantum-cost}). Frequent rotation of Session Ticket Encryption Keys (STEKs) severs the resumption chain entirely.

\subsection{Multiplying Quantum Friction: Rekeying}
\label{sec:mit-rekey}

Rekeying targets the quantum axis. Each independent DH exchange forces a fresh run of Shor's algorithm with negligible storage impact ($\approx$2--3\,KB per rekey). This mechanism elevates the scarcest adversary resource: quantum computation time. As established in Section~\ref{sec:cost-analysis}, for a session of length $P$, if rekeying occurs every $R$~bytes, the adversary must solve $E$ independent instances of the discrete logarithm problem. For full transcript recovery, this instance multiplier is $E = \lceil P/R \rceil$.

\begin{figure}[t]
\centering
\includegraphics[width=\columnwidth]{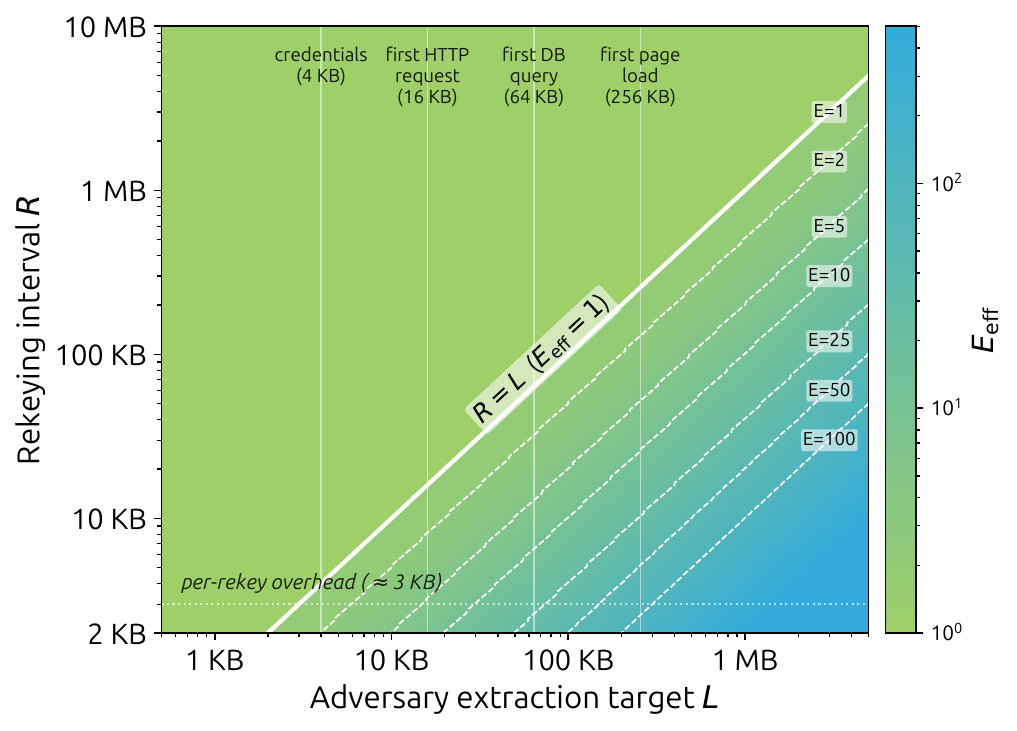}
\caption{Effective quantum multiplier $E_\mathrm{eff}(L,R) = \lceil L/R \rceil$. $R$ denotes the byte budget between DH exchanges. Above the $R = L$ diagonal, $E_\mathrm{eff} = 1$: rekeying provides zero protection against targeted prefix extraction. The dotted line marks the per-rekey overhead floor, where connection bandwidth becomes overwhelmingly dominated by cryptography. Vertical guides indicate ballpark extraction targets ranging from a credential exchange ($\sim$4\,KB) to a full initial page load ($\sim$256\,KB), representing order-of-magnitude application-layer payload sizes after the handshake.}
\label{fig:partial-E}
\end{figure}

This generalizes to a \emph{partial-decryption adversary} requiring only $L \leq P$ bytes of target plaintext (e.g., initial authentication tokens). The effective quantum multiplier becomes
\begin{equation}
\label{eq:E-eff}
E_\mathrm{eff}(L, R) \;=\; \left\lceil \frac{\min(L, P)}{R} \right\rceil,
\end{equation}
which collapses to $E_\mathrm{eff} = 1$ whenever $R \geq L$. Figure~\ref{fig:partial-E} maps this dynamic. To protect credential-style targets ($L \sim \text{KB}$), the threshold $R$ must be driven down to $R \lesssim L$, a regime where handshake overhead rivals the payload itself.

\subsubsection{SSH RekeyLimit}
\label{sec:mit-ssh-rekey}

SSH supports true in-band ephemeral rekeying: a \texttt{SSH\_MSG\_KEXINIT} triggers a full Diffie-Hellman exchange within the existing TCP connection, producing a fresh shared secret that constitutes an independent quantum problem. The configurable \texttt{RekeyLimit} parameter defines a nominal plaintext threshold ($R_\text{nom}$). However, OpenSSH's internal implementation bounds this limit based on the total size of encrypted transport packets transmitted (which includes headers, MACs, and padding), rather than bare application payload. Because typical SSH payloads are highly fragmented, each transport packet is padded heavily before encryption. Consequently, the application payload actually transmitted before a rekey triggers is significantly larger than the configured limit; in practice, $R_\text{nom} = 64$~KB results in an effective payload threshold of $R_\text{eff} \approx 127$~KB. This dynamic is empirically validated using the testbed (Section~\ref{sec:exp-testbed}), sweeping four \texttt{RekeyLimit} parameters (64~KB--10~MB) across five payloads (1~KB--5~MB, 25 captures).

\begin{figure}[htbp]
\centering
\includegraphics[width=\columnwidth]{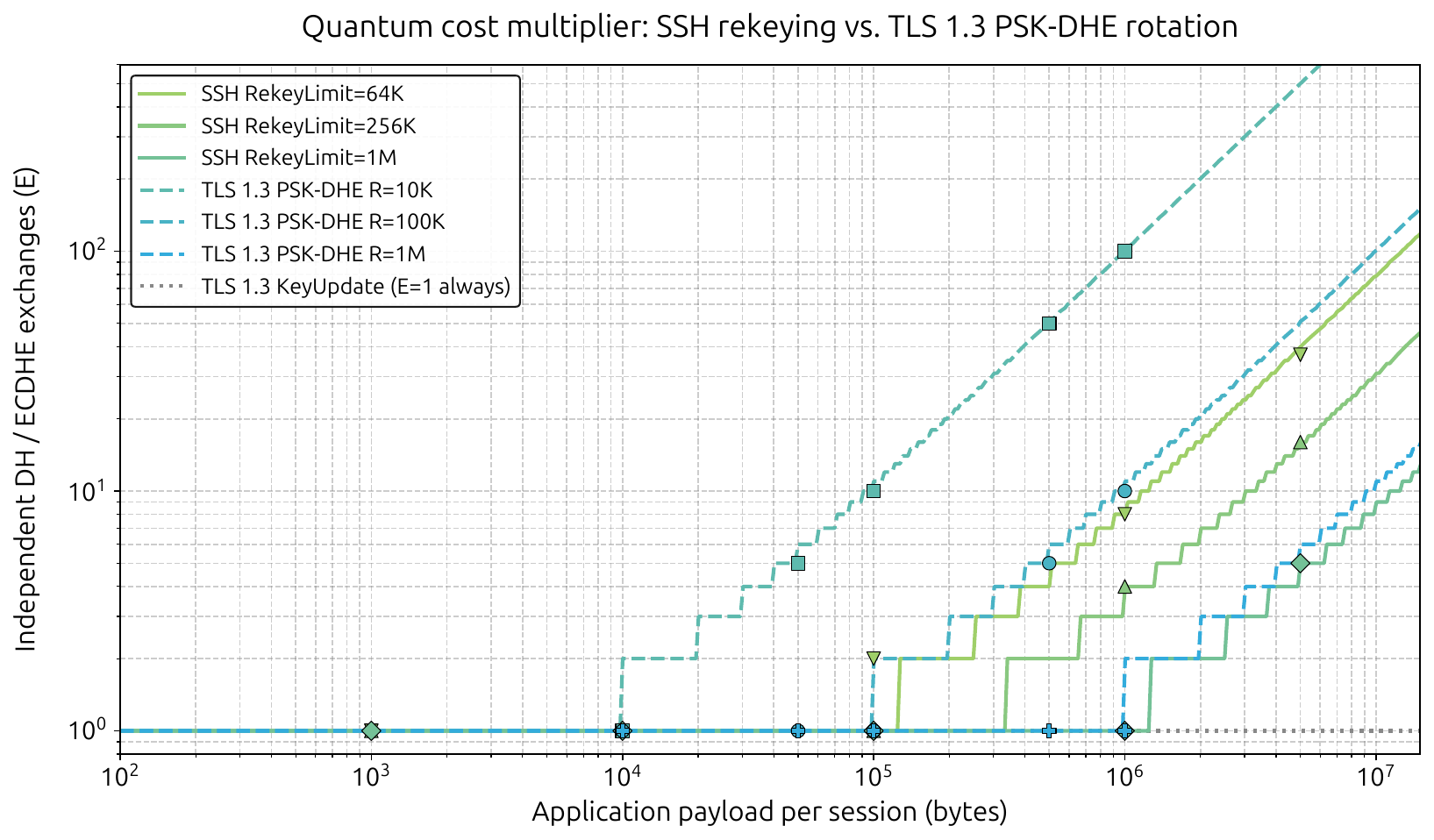}
\caption{Quantum cost multiplier~$E$ for SSH rekeying and TLS~1.3 PSK-DHE rotation. Markers denote measured values from the testbed; curves show $\lceil P/R_\mathrm{eff}\rceil$. SSH uses effective thresholds derived from transport-level byte counting ($R_\mathrm{eff} \approx 2 R_\mathrm{nom}$). TLS markers confirm $E = \lceil P/R\rceil$ exactly (15 captures, all match).}
\label{fig:rekey-E}
\end{figure}

Figure~\ref{fig:rekey-E} demonstrates this asymmetric efficiency. At a 5~MB payload, the most aggressive limit ($R = 64$~KB, $E = 37$) imposes only a 2.1\% storage penalty~$\alpha$, while the quantum instance multiplier~$E$ scales linearly.

\subsubsection{Quantum Cost Implications}
\label{sec:mit-quantum-cost}

The multiplier~$E_\mathrm{eff}$ scales aggregate quantum workload regardless of the per-instance computational cost~$T_q$. Solving 256-bit ECDLP currently demands ${\sim}10^{11}$ Toffoli gates across ${\sim}2{,}330$ logical qubits, placing $T_q$ in the hours-to-days regime~\cite{roetteler2017quantum,haner2020improved}. Selecting larger group parameters inflates $T_q$ independently: Shor's gate complexity grows superlinearly with key size, so migrating from P-256 to P-384 increases per-instance quantum cost with no measurable bandwidth penalty. The total quantum cost per session, $E \times T_q$, is therefore inflatable along two orthogonal and composable axes: rekeying frequency ($E$) and key exchange parameter size ($T_q$). Even assuming an aggressive future $T_q = 1$~h on the current P-256 curve, enforcing an SSH rekey interval of 64~KB ($E = 37$) extends full transcript decryption latency from an hour to nearly two days. 

TLS~1.3 has no equivalent mid-stream mechanism. Its \texttt{KeyUpdate} message derives each new traffic secret from its predecessor via a single HKDF expansion with no fresh randomness (RFC~8446~\S7.2); recovering the initial handshake secret therefore exposes every subsequent epoch, leaving $E=1$ regardless of session length. The only alternative is PSK-DHE session resumption, where each resumed connection performs a fresh ECDHE exchange. Reusing the testbed (Section~\ref{sec:exp-testbed}), per-resumption overhead was measured over $n=10$ chains ($H_\text{init} = 3{,}180$~B, $H_\text{resum} = 2{,}034$~B, zero variance), and 15 data-transfer captures across three intervals ($R \in \{10\text{K}, 100\text{K}, 1\text{M}\}$~B) confirmed $E = \lceil P/R \rceil$ exactly (Figure~\ref{fig:rekey-E}). PSK-DHE rotation thus achieves the same cryptographic scaling as SSH rekeying. The critical difference is transport cost: SSH injects fresh key material inside an existing TCP connection in a single round trip, whereas TLS~1.3 requires establishing a new TLS connection (and therefore a new TCP connection and handshake) for every interval. At aggressive rekey thresholds this latency penalty is prohibitive, making rapid rotation a theoretical bound rather than a viable engineering solution. QUIC inherits the same limitation; its key update (RFC~9001~\S6) uses an identical deterministic HKDF chain. The Extended Key Update drafts for TLS~\cite{ietf-tls-extended-key-update-09} and QUIC~\cite{ietf-quic-extended-key-update-02} eliminate this limitation by injecting a fresh Diffie-Hellman exchange into the live session, re-seeding the key schedule with independent entropy and restoring the $E$-scaling that SSH already provides.

\subsection{The Limits of Storage Inflation: Record Padding}
\label{sec:mit-padding}

Because raw ciphertext dictates storage volume, inflating protocol overhead~$\alpha$ mathematically increases the adversary's burden. RFC~8446 permits padding TLS~1.3 records with zero bytes before AEAD encryption. IND-CPA security forces the harvester to archive every byte.

For block size~$b > 0$, a record of plaintext length~$p$ bytes pads to the next multiple of~$b$. The encrypted wire length is:
\begin{equation}
\label{eq:padded-record}
\bigl\lceil (p + 1) / b \bigr\rceil \cdot b + r + t \quad\text{bytes,}
\end{equation}
where $r = 5$~B is the header and $t = 16$~B the AEAD tag. 

\begin{figure}[htbp]
\centering
\includegraphics[width=\columnwidth]{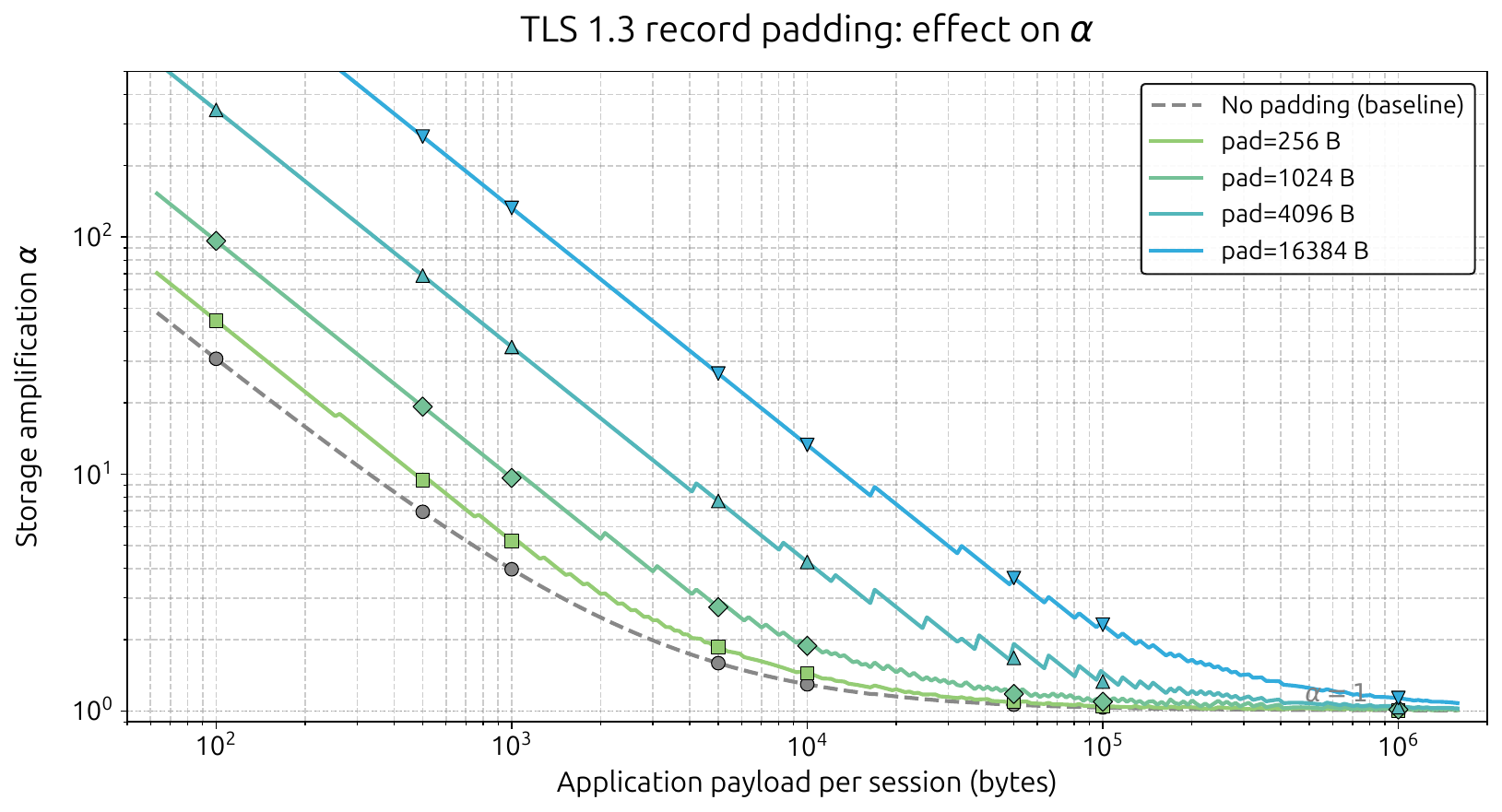}
\caption{Maximum theoretical padding ($b = 16{,}384$~B) aggressively inflates~$\alpha$ for small sessions but imposes negligible friction on large transfers.}
\label{fig:padding-alpha}
\end{figure}

This model is empirically validated using the testbed (Section~\ref{sec:exp-testbed}), sweeping five padding block sizes ($b \in \{0, 256, 1\text{K}, 4\text{K}, 16\text{K}\}$~B) across eight payloads (100~B--1~MB, 40 captures). As shown in Figure~\ref{fig:padding-alpha}, for 100~B sessions, maximum padding inflates~$\alpha$ from $30\times$ to $1{,}300\times$. 

Massive bandwidth padding is economically mismatched against a state-actor threat model. Every padded byte must traverse real network infrastructure: the defender pays active bandwidth and egress costs in proportion to the inflation, while the state adversary absorbs the volume passively via optical tape arrays. This structural asymmetry is inherent to the~$\alpha$ axis; storage inflation penalizes both sides, favoring the party with lower marginal cost per byte. Quantum-axis inflation ($E$, $T_q$) does not share this limitation. Padding therefore serves primarily as an academic upper bound, while rekeying and key-size selection offer more cost-effective defensive leverage.

%% ====================================================================
%%  Section 7 — Discussion
%% ====================================================================
\section{Discussion}
\label{sec:discussion}

NIST finalized its primary post-quantum key encapsulation standard (ML-KEM) in 2024. Hybrid key exchanges combining X25519 with ML-KEM-768 are already active in production browsers and global CDNs~\cite{nist2024fips203,cloudflare2025pq}. Simultaneously, the NSA mandates post-quantum capability for national security systems by 2025, culminating in exclusive use by 2035~\cite{nsa2022cnsa2}.

For TLS~1.3, this migration is straightforward: hybrid algorithms integrate directly into the existing \texttt{supported\_groups} extension. Similarly, hybrid post-quantum key exchange for SSH is under active standardization and experimentation~\cite{kampanakis2020pqssh}.

The operational bottleneck is \emph{crypto-agility}: the capability to rotate algorithms swiftly upon cryptanalytic compromise. Protocols that hard-code algorithms or require firmware updates (e.g., IoT, SCADA) cannot pivot within relevant timescales. For sectors where data retention spans decades (e.g., 25--50~years in healthcare~\cite{bolaji2023cost}), Mosca's inequality dictates that migration deadlines have already passed. Forward secrecy's limits are absolute: once encrypted traffic transmits, the sender loses cryptographic control over its lifespan. PQC migration arrests future accumulation but cannot rewrite the past.

This irreversibility mandates a defensive triage. Data with confidentiality requirements exceeding the PQC horizon (state secrets, genome records) demands immediate migration. Data whose sensitivity decays rapidly (session cookies, ephemeral messaging) may tolerate delay. The models in Section~\ref{sec:cost-analysis} govern this prioritization: $\alpha$ quantifies the storage resource the adversary has already committed, while $E$ projects the required quantum effort for future decryption.

The measures evaluated in Section~\ref{sec:mitigation} do not cure HN-DL; every classical key exchange remains fundamentally broken by Shor's algorithm. Their value is defense in depth along the adversary's cost axes. As Section~\ref{sec:mitigation} establishes, the quantum axis ($E \times T_q$) is inherently adversary-facing, while storage inflation ($\alpha$) penalizes both sides. The quantum axis is therefore the superior defensive degree of freedom, though $E_\mathrm{eff}$ collapses when the adversary targets only prefix secrets.

A critical protocol asymmetry constrains this strategy. SSH supports in-band ephemeral rekeying mid-stream ($\approx$2--3\,KB overhead per rekey), but TLS~1.3 (over 93\% of connections~\cite{cloudflare2024pq}) and QUIC lack an equivalent deployed mechanism: \texttt{KeyUpdate} is deterministic ($E=1$), and achieving $E > 1$ requires a new connection (Section~\ref{sec:mit-quantum-cost}). The Extended Key Update drafts~\cite{ietf-tls-extended-key-update-09,ietf-quic-extended-key-update-02} would remove this constraint, but they have not yet reached production; the quantum cost axis therefore remains unreachable for the protocols that carry the majority of Internet traffic.

Our storage taxonomy assumes a passive network adversary with sustained access to full sessions. Constrained collection strategies (metadata-only retention or statistical packet sampling) face different economic realities decoupled from~$\alpha$. Active endpoint compromise bypasses the HN-DL model entirely. The mechanism by which the adversary first gains access to encrypted traffic, whether through backbone taps, ISP cooperation, or data-center presence, falls outside our scope but remains an operationally significant prerequisite~\cite{pclob2023section702,prg2013liberty}.

Our cost model deliberately adopts the adversary's worst case. The \$12.16/TB-year cloud reference price is a fully loaded operational figure that bundles hardware amortization, personnel, electricity, cooling, and geographic redundancy. A state actor operating dedicated tape libraries pays none of these recurring charges; the dominant expense is a one-time media CapEx (\$5.25/TB for LTO-9~\cite{HPE_LTO9_2024}) whose marginal retention cost reduces to floor space and minimal power. Recent AI-driven demand has inflated DRAM and NAND flash prices, with secondary pressure on HDD procurement~\cite{reuters2025memorychip}. Cloud providers absorb and pass through these spikes, widening the $\delta < 0$ tail in our Monte Carlo model. An adversary whose archival hardware is already deployed does not see these price movements; the investment is already paid. Our estimates therefore represent an upper envelope; the true retention cost for a dedicated state program is likely lower and substantially more stable than any scenario our simulation samples.

%% ====================================================================
%%  Section 8 — Conclusion
%% ====================================================================
\section{Conclusion}
\label{sec:conclusion}

Harvest-now, decrypt-later is not a speculative risk: national security agencies warn that adversaries may already be stockpiling ciphertext for future quantum decryption~\cite{anssi2024qkd,NIST-hndl}. This paper quantifies HN-DL feasibility across TLS~1.2, TLS~1.3, QUIC, and SSH, demonstrating that retro-decryption is perfectly deterministic once the ephemeral key collapses.

Our overriding finding is economic: once an adversary can access encrypted traffic, retaining it is trivial. Even under a conservative, fully loaded commercial cloud upper bound, storing a 1\% global harvest costs only $\mathcal{O}(10^9)$~USD per year; dedicated tape archives push the raw media cost lower still. Monte Carlo projection over a 10-year retention horizon confirms this scale even under a 10\% annual price increase, itself a cloud-only ceiling that overstates costs for state actors with sunk infrastructure CapEx, placing cumulative expenditure at $\mathcal{O}(10^{10}$--$10^{11})$~USD. Gaining that access, whether through backbone fiber taps, ISP mandates, or data-center presence, remains a separate and operationally significant prerequisite that this work deliberately excludes.

The adversary's cost map exposes two defensive axes with fundamentally different economics. Inflating~$\alpha$ (record padding) penalizes both sides proportionally; every additional byte must traverse real network infrastructure, making storage-axis inflation structurally self-harming. Inflating the quantum axis $E \times T_q$ is inherently adversary-facing: aggressive rekeying multiplies required Shor instances ($E_\mathrm{eff} = \lceil L/R \rceil$), and larger key exchange parameters inflate per-instance cost ($T_q$), both with negligible wire overhead. ECH further degrades metadata triage, forcing affordable selective harvesting into unaffordable bulk collection. Administrators must concurrently deprecate non-forward-secret modes (TLS~1.2 RSA, TLS~1.3 0-RTT) that expose entire session histories to a single key compromise.

Permanent defense requires completing the migration toward post-quantum key encapsulation (ML-KEM). The accumulated archive of pre-quantum ciphertext remains an unavoidable liability, and delay in transition expands this exposure daily. The full simulation pipeline and experimental artifacts are released as open-source software~\cite{perlabuc3m_hndl}.

Several open problems remain. TLS~1.3 and QUIC remain locked at $E=1$; the Extended Key Update drafts~\cite{ietf-tls-extended-key-update-09,ietf-quic-extended-key-update-02} would unlock the quantum cost axis, but deployment has not begun. Our storage asymptote $\alpha$ was validated via loopback captures; real-network MTU fragmentation remains a natural extension. The economics of partial PQC deployment also deserve study: because TLS and SSH negotiate key exchange parameters in cleartext, an adversary can discard quantum-resistant sessions and concentrate harvesting on the shrinking classical remainder, an efficiency gain that current triage models do not capture.

%% ====================================================================
%%  Appendices
%% ====================================================================
\appendix
\section{TLS 1.3 Derivation Details}
\label{appendix:tls13-singlehndl}

Once the ECDHE shared secret~$Z$ is known, the TLS~1.3 key schedule~\cite{rfc8446,rfc5869} is a deterministic chain: every handshake and application traffic secret follows without additional randomness, computable from the captured transcript alone. This appendix documents two transcript hash pitfalls encountered during offline reconstruction of the key schedule from captured traffic. Both produce silently incorrect keys: HKDF returns output regardless of input correctness, so the error surfaces only when verification against ground truth fails.

\subsection{Session Resumption and PSK Cascade}
\label{appendix:tls13-resumption}

After a full handshake, the server issues \texttt{NewSessionTicket} messages. Each ticket carries a \texttt{ticket\_nonce} bound to a PSK derived from the Resumption Master Secret~\cite{rfc8446}.

Correct offline reconstruction requires careful attention to the transcript hash boundary. The application traffic secrets bind to $\text{Hash}(\mathit{CH}\cdots\mathit{SF})$, the transcript through Server Finished. The \texttt{res master} label, however, extends one message further: it binds to $\text{Hash}(\mathit{CH}\cdots\mathit{CF})$, including Client Finished. In a live TLS stack this distinction is invisible because the running hash accumulates messages as they arrive. In offline reconstruction from a PCAP, both hashes are computed manually; reusing the shorter form for \texttt{res master} yields a wrong Resumption Master Secret and silently breaks decryption of every resumed session in the PSK chain.

The HN-DL implication is a cascade: breaking one session's ECDHE exchange recovers its MasterSecret, its Resumption Master Secret, and every PSK derived from it. Any resumed session that uses one of those PSKs in pure-PSK mode becomes retroactively decryptable. PSK-DHE mode is not affected: the \texttt{HandshakeSecret} incorporates a fresh ephemeral \texttt{(EC)DHE} shared secret, so the adversary must also solve the discrete logarithm for the resumed session's ephemeral keys.

\subsection{0-RTT Early Data}
\label{appendix:tls13-0rtt}

0-RTT early data is encrypted under keys derived solely from a prior session's PSK, without a fresh ECDHE exchange~\cite{rfc8446}. Because no fresh ephemeral secret enters the early key schedule, the confidentiality of 0-RTT data is retroactively broken by a single ECDHE compromise anywhere in the PSK chain; this is inherent to the protocol design (RFC~8446, \S E.1.3).

A second transcript hash pitfall arises here. The PSK binder is a MAC proving knowledge of the PSK, so it cannot include itself in its own input; RFC~8446 \S4.2.11.2 therefore computes it over a \emph{truncated} ClientHello that excludes the binders list. The early traffic secret, however, uses the \emph{full} ClientHello including the binders (RFC~8446 \S7.1). An implementer who reuses the truncated form for traffic key derivation obtains silently incorrect early traffic keys.

\section{Minimal-Archive Analysis}
\label{appendix:minimal-archive}

An adversary could reduce the data-phase overhead~$\omega$ by stripping per-record fields not required for retrospective decryption.

\paragraph{AEAD tags.}
Authentication tags are not inputs to the decryption function. AES-GCM decrypts via AES-CTR on the ciphertext body independently of the GHASH tag~\cite{rfc5116}, and ChaCha20 produces its keystream independently of the Poly1305 tag~\cite{rfc8439}. All 16-byte tags are therefore unconditionally strippable. Beyond tags, each protocol imposes structural constraints that prevent further reduction.

\paragraph{TLS 1.3.}
Each record carries a five-byte cleartext header followed by the AEAD ciphertext~\cite{rfc8446}. The first three bytes are constant (\texttt{0x17\,03\,03}) and reconstructible; the two-byte \texttt{length} field is needed for record boundaries. The nonce is implicit (sequence number $\oplus$ \texttt{write\_iv}; RFC~8446, \S5.3), and the inner content type is encrypted inside the payload. Result: $\omega_{\min} \leq 3$\,B vs.\ $\omega = 22$\,B.

\paragraph{SSH (chacha20-poly1305).}
The Binary Packet Protocol~\cite{rfc4253} frames each packet with a 4-byte \texttt{packet\_length} encrypted under~$K_1$ (needed as boundary delimiter), followed by \texttt{padding\_length}, payload, and random padding (averaging 7.5\,B for 8-byte alignment), all encrypted under~$K_2$. Padding and payload are indistinguishable without decryption, so the entire ciphertext must be retained; only the 16-byte Poly1305 tag is strippable. Result: $\omega_{\min} \approx 12.5$\,B vs.\ $\omega \approx 28.5$\,B.

\paragraph{QUIC.}
1-RTT short-header packets~\cite{rfc9000} contain a header-protected flags byte (Key Phase, Packet Number Length), a Destination Connection ID (typically 8\,B, needed for demultiplexing), and a Packet Number (typically 2\,B, needed for the AEAD nonce; RFC~9001, \S5.3). All three are essential; only the 16-byte tag is strippable. Result: $\omega_{\min} = 11$\,B vs.\ $\omega = 27$\,B.

\paragraph{Impact.}
Table~\ref{tab:minimal-omega} summarises the achievable savings. The maximum difference $|\alpha_\infty - \alpha_{\infty,\min}|$ is $1.2 \times 10^{-2}$ (QUIC) and below $1.2 \times 10^{-3}$ for TCP-based protocols, confirming that per-record stripping yields no strategically meaningful storage advantage.

\begin{table}[htbp]
\centering
\caption{Per-record overhead: full capture vs.\ minimal archive. $\alpha_\infty = 1 + \omega/M$.}
\label{tab:minimal-omega}
\small
\begin{tabular}{@{}lrrrr@{}}
\hline
\textbf{Protocol} & $\omega$ & $\omega_{\min}$ & $\alpha_\infty$ & $\alpha_{\infty,\min}$ \\
\hline
TLS 1.3 &  22\,B &  3\,B & 1.0013 & 1.0002 \\
SSH     & 28.5\,B & 12.5\,B & 1.0009 & 1.0004 \\
QUIC    &  27\,B &  11\,B & 1.020  & 1.008  \\
\hline
\end{tabular}
\end{table}

\section*{Acknowledgment}
This work was supported by the Spanish Government through the grant DIstributed Smart Communications with Verifiable Energy-optimal Yields (DISCOVERY), PID2023-148716OB-C33, funded by MICIU/AEI/10.13039/501100011033. It was also supported by the Comunidad de Madrid through the grant ``RAMONES-CM'', TEC-2024/COM-504.

\printbibliography

\end{document}